# Time-Dependent Ionization in the Envelopes of Type II Supernovae at the Photospheric Phase

M. Sh. Potashov[1*], S. I. Blinnikov[1,2,3**], and V. P. Utrobin[1***]

[1]*Institute for Theoretical and Experimental Physics, ul. Bol'shaya Cheremushkinskaya 25, Moscow, 117218 Russia*
[2]*Sternberg Astronomical Institute, Moscow State University, Universitetskii pr. 13, Moscow, 119992 Russia*
[3]*Institute of Physics and Mathematics of the Universe, University of Tokyo, Kashiwa, Japan*

**Abstract**—The importance of allowance for the time-dependent effect in the kinetics at the photospheric phase during a supernova explosion has been confirmed by several independent research groups. The time-dependent effect provides a higher degree of hydrogen ionization in comparison with the steady-state solutions and strengthens the H$\alpha$ line in the resulting simulated spectrum, with the intensity of the effect increasing with time. However, some researchers argue that the time-dependent ionization effect is unimportant. Its allowance leads to an insignificant strengthening of H$\alpha$ in their modeling only in the first days after explosion. We have demonstrated the importance of the time-dependent effect with the models of SN 1999em as an example using the new original LEVELS software package. The role of a number of factors that can weaken the time-dependent effect has been checked. We have confirmed that the intensity of the effect is affected by the abundance of metal admixtures in the envelope, while the addition of extra levels to the model hydrogen atom weakens the time-dependent effect to a lesser degree and never removes it completely.

## INTRODUCTION

New data, the photometric distances to objects with known redshifts, are required to investigate the present-day structure of the Universe. Among the great variety of distance measurement techniques there are methods that do not rely on the cosmological distance ladder, for example, the expanding photosphere method (EPM) (Kirshner and Kwan 1974), the spectral-fitting expanding atmosphere method (SEAM) (Baron et al. 2004), or the dense shell method (DSM) (Blinnikov et al. 2012; Potashov et al. 2013; Baklanov et al. 2013) that use type IIP and IIn supernovae (SNe) as objects. Note that using such a method as the SEAM requires the construction of a complete physical model for a type II SN that reproduces in detail its spectrum.

To completely model the physical processes occurring in a SN, it is necessary to simultaneously take into account the envelope expansion hydrodynamics, the matter–radiation field interaction, the radiative transfer in lines and continuum, and the kinetics of level populations in the atoms of a multiply charged plasma. This gives a system of partial integro-differential equations of radiation hydrodynamics that cannot yet be completely solved numerically even in the one-dimensional case. One has to resort to unavoidable simplifications in this complete system. One of such simplifications is the steady-state approximation of the kinetic system of level populations, when the system is assumed to be in statistical equilibrium.

The importance of allowance for the time-dependent effect in the kinetics at the nebular phase for SN II was pointed out by Axelrod (1980), Clayton et al. (1992), and Fransson and Kozma (1993). It was shown that several years after explosion (∼800 days for SN 1987A; Fransson and Kozma 1993; Kozma and Fransson 1998a, 1998b; Jerkstrand et al. 2011) hydrogen in the SN envelope begins to experience the time-dependent effect. Allowance for this effect led to an increase in the degree of ionization and matter temperature by a factor of 2–4 and to an enhancement of the emission compared to the steady-state approximation. It is important to note that in these papers the time-dependent effect was taken into account not only in the kinetic and ionization equations but also in the energy equation.

[*]E-mail: Marat.Potashov@gmail.com
[**]E-mail: Sergei.Blinnikov@itep.ru
[***]E-mail: Utrobin@itep.ru

The time-dependent hydrogen ionization effect in the envelopes of type II SN at the photospheric phase was used by Kirshner and Kwan (1975) to explain the high H$\alpha$ intensity in the spectra of SN 1970G and by Chugai (1991) to explain the high degree of hydrogen excitation in the outer atmospheric layers ($v > 7000$ km s$^{-1}$) of SN 1987A in the first 40 days after explosion.

Utrobin and Chugai (2002) found a strong time-dependent effect in the ionization kinetics and hydrogen lines in type IIP SN during the photospheric phase. In their next paper Utrobin and Chugai (2005b) also took into account the time-dependent effect in the energy equation. Initially, independent hydrodynamic modeling of the envelope was performed, and, subsequently, the time-dependent equation for the matter temperature and the complete kinetic system of level populations for both atoms and molecules were solved with the available matter density profile, expansion velocities, photospheric radius, and effective temperature. If we replace the time-dependent energy equation by the energy balance equation (the steady-state approximation) and the kinetic system by the statistical equilibrium equations, then the simultaneous solution of this new system will be exactly the steady-state approximation in Utrobin and Chugai (2005b). An important consequence of these papers was the conclusion that including the time-dependent ionization allowed the spectra of peculiar SN 1987A with a stronger H$\alpha$ line to be obtained, which could not be done previously without mixing radioactive $^{56}$Ni into the outer high-velocity layers in the steady-state approximation. In the next paper (Utrobin 2007) the importance of this effect was also shown for normal SN 1999em.

The conclusions reached by Utrobin and Chugai were confirmed by Dessart and Hillier using the `CMFGEN` software package. In Dessart et al. (2008) the applied approach was still the steady-state one, which was implemented in the `CMFGEN` package. Modeling revealed a problem: the H$\alpha$ line in hydrogen-rich envelopes was weaker than that observed at the recombination epoch. In particular, the model did not reproduce the line for times later than four days for SN 1987A and later than 20 days for SN 1999em. Next, Dessart and Hillier improved the code by including the time dependence in the kinetic system and the energy equation (Dessart and Hillier 2007) and then in the radiative transfer (Dessart and Hillier 2010; Hillier and Dessart 2012). This allowed the H$\alpha$ line to be strengthened in the resulting simulated spectrum at better agreement with observations. It can be added that the density profile and the abundances of elements for `CMFGEN` were taken from independent hydrodynamic simulations with the `KEPLER` code (for more details see Hillier and Dessart 2012).

On the other hand, based on their computations with the `PHOENIX` software package, De et al. (2010a) found the time-dependent kinetics to be important only in the first days after SN explosion. Moreover, they argue that the role of the time-dependent effect is not very strong even in these first days by illustrating this with the models of SN 1987A and SN 1999em as an example.

Thus, the conclusions of the various research groups disagree. Within our formulation of the problem we will attempt to answer the question of whether the time-dependent ionization effect is important or not and if it is important, then what can affect its intensity.

The paper consists of three parts. First, we will describe the details of our method of modeling the SN envelope physics. Then, within our model we will illustrate the time-dependent effect and discuss the criteria for a deviation from the steady-state approximation. In the last section we will talk about the physical processes affecting the effect of time dependence in the kinetics under consideration.

## MODELING

We take the profiles of density $\rho(r,t)$, envelope expansion velocity $v(r,t)$, matter temperature $T_e(r,t)$, radiation color temperature $T_c(t)$, and the photospheric radius $R_{\rm ph}(t)$ from our modeling with the `STELLA` code (Blinnikov et al. 1998, 2000, 2006). Initially, the presupernova is modeled. In the next step, the opacity table is computed by taking into account the absorption in spectral lines in a medium with a velocity gradient (the expansion opacity). Finally, the time-dependent radiative transfer equation is solved for each frequency group in the two-moment approximation in each Lagrangian zone simultaneously with the hydrodynamic equations. The equation of state treats the ionization in Saha's equilibrium approximation. The code also takes into account the scattering of photons by electrons.

The hydrodynamic and thermodynamic parameters of the supernova obtained with `STELLA` are subsequently used as input into `LEVELS` for a self-consistent solution of the complete time-dependent (or steady-state) kinetic system of level population equations with the radiative transfer equations in Sobolev's approximation when the LTE approximation is already abandoned. In this step the chemical composition of the expanding envelope can be specified. In this paper we will be interested in the models with a purely hydrogen envelope. An example of considering purely hydrogen SN envelopes is encountered in the literature (Duschinger et al. 1995). Such an artificial simplification turns the original problem into an academic one but will allow the time-dependent effect to be investigated in detail.

To study the time-dependent effect, we take SN 1999em with a well-defined plateau phase in its light curve. We consider the R450_M15_Ni004_E7 model from Baklanov et al. (2005) with a presupernova radius $R = 450 R_\odot$, mass $M = 15 M_\odot$, and an explosion energy of $7 \times 10^{50}$ erg at a distance $D = 7.5$ Mpc to the galaxy NGC 1637, where the supernova exploded, despite the fact that a more correct distance estimate based on a lot of data is $D \approx 12$ Mpc. For the purposes of this paper the question about the accuracy of the distances to the host galaxy is unimportant. We consider the mentioned model as a typical one for type II SN.

### *Continuum Radiation*

Basically, STELLA provides the intensity of continuum radiation $J_\nu(t,r)$ averaged over the angles and over a group of frequencies $\nu$ for each time $t$ at radius $r$. Therefore, there is no need for temperature parametrization of the radiation field. However, in the current formulation of the problem $J$ is specified in a simplified form, just as was done by Utrobin and Chugai (2005). Such a description of the radiation fields will allow us to solve the system of kinetic equations simultaneously with the transfer equation in lines for each Lagrangian mass zone independently.

The optical depth for photons of the Lyman continuum $LyC$ of neutral hydrogen is very large. Therefore, there is virtually no photospheric radiation in the frequency band $\nu \geqslant \nu_{LyC}$, and the radiation field in this case is determined for the regions above the photosphere by diffusive radiation. Each such hard photon emitted upon direct recombination is immediately absorbed. The transfer equation is simplified for an optically thick medium:

$$J_\nu(t,r) = S_\nu = \eta_\nu/\kappa_\nu, \qquad (1)$$

where $S_\nu$ is the source function, $\eta_\nu$ and $\kappa_\nu$ are the emission coefficient and the true absorption coefficient corrected for stimulated emission, respectively (Mihalas 1978; Hubeny and Mihalas 2014). In this frequency band the free–bound processes (Fig. 1) make a major contribution to the opacity, and Eq. (1) can be written as $S_\nu = \eta_\nu^{fb}/\kappa_\nu^{bf}$. In this case, the photoionization rate from the ground hydrogen level and the recombination rate to the ground level closely coincide. We will use this approximation. It should be noted that at the frequencies under consideration the relatively small contributions of the lines and free–free processes to the emission and absorption coefficients will upset the detailed balance. However, our simplified consideration does not "switch off" the time-dependent ionization effect (Utrobin and Chugai 2005a).

For the ultraviolet and optical bands, $\nu < \nu_{LyC}$, we use the approximation of an optically thin medium. Therefore, the mean intensity can be written as

$$J_\nu(t,r) = W B_\nu(T_c), \qquad (2)$$

where $W$ is the geometrical dilution factor, $B_\nu$ is the intensity of blackbody radiation, and $T_c$ is the radiation color temperature determined by fitting the photospheric spectrum of the STELLA model. It is important to note that for frequencies between the Lyman and Balmer ionization thresholds, because of the great expansion opacity (Fig. 1) provided mostly by numerous metal lines, the averaged intensity of such a quasi-continuum will be lower than that in the optically thin limit even if we take into account the fact that 71% of the two-photon emission lies between these thresholds (Nussbaumer and Schmutz 1984; Xu et al. 1992). However, in our analysis, for the time being, we dwell on an approximate case to illustrate the time-dependent ionization.

### *Radiative Transfer in Lines*

Our STELLA modeling shows that the transition to a homologous (with a high accuracy) expansion of the SN 1999em envelope ends approximately by day 15 after explosion (Baklanov et al. 2005). This fact simplifies the application of the Sobolev method (Sobolev 1960; Castor 1970) to describe the radiative transfer in lines.

In the current formulation of the problem we use the classical Sobolev approximation in which the mean intensity of radiation in an $l \to u$ transition averaged over the line profile is

$$J_{lu} = (1 - \beta_{lu})S_{lu} + \beta_{lu}J(\nu_{lu}), \qquad (3)$$

where

$$\beta_{lu} = \frac{1 - \exp(-\tau_{lu})}{\tau_{lu}} \qquad (4)$$

is the photon escape probability integrated over the directions and over the line frequencies, and

$$S_{lu}(r) = \frac{2h\nu_{lu}^3}{c^2}\left(\frac{g_u n_l}{g_l n_u} - 1\right)^{-1} \qquad (5)$$

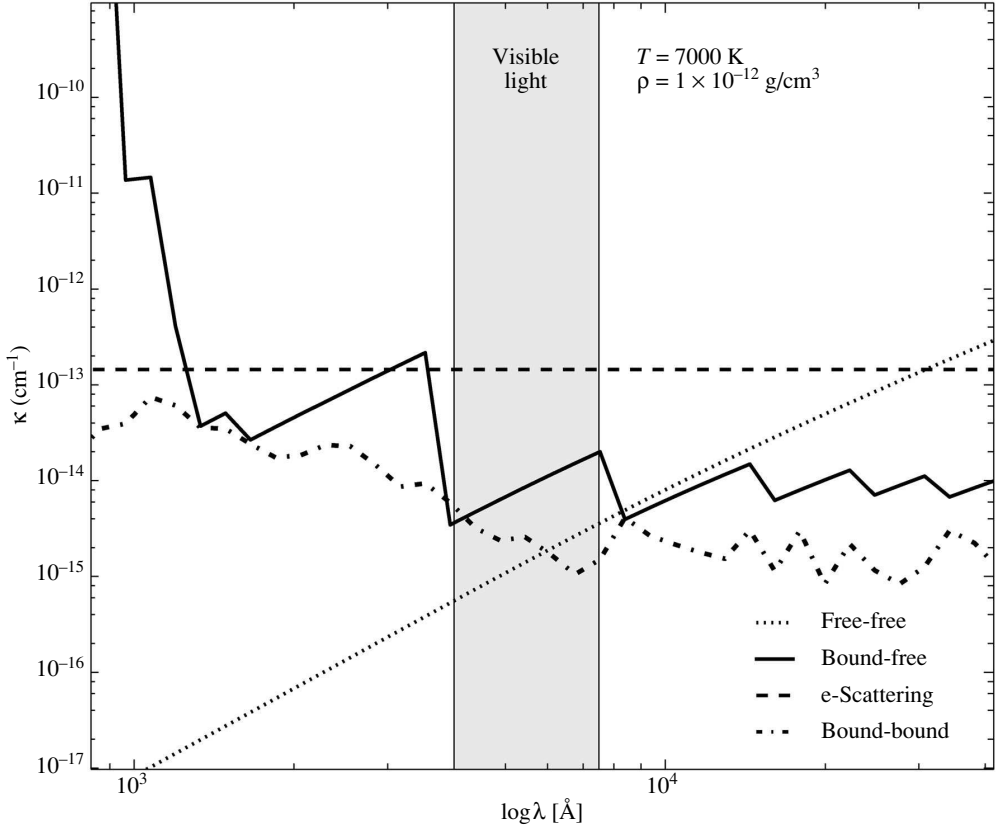

**Fig. 1.** Coefficients of extinction due to various processes: electron scattering, free−free absorption, free−bound absorption, bound−bound absorption in spectral lines in a medium with a velocity gradient (expansion opacity). From the `STELLA` computations.

is the source function; $\nu_{lu}$ is the line frequency;

$$\tau_{lu} = \frac{c^3}{8\pi} \frac{1}{\nu_{lu}^3} \frac{g_u}{g_l} A_{ul} t \left( n_l - \frac{g_l}{g_u} n_u \right) \quad (6)$$

is the Sobolev optical depth in the $l \to u$ transition. Here, $A_{ul}$ is the Einstein coefficient for the spontaneous $u \to l$ transition; $n_l$, $g_l$, $n_u$, and $g_u$ are the populations and statistical weights of the atom at the lower and upper levels, respectively.

It is important to note that in the Sobolev approximation the ordinary spontaneous transition rate $A_{ul}$ is replaced by the effective one $A_{ul}^{\text{eff}} = A_{ul} \beta_{lu}$, which is the rate of uncompensated radiative transitions. If, in addition, we take into account the fact that $n_l \gg n_u$, then in the case of a large Sobolev optical depth, when $\beta_{lu} \sim 1/\tau_{lu}$, we obtain

$$A_{ul}^{\text{eff}} \sim \frac{8\pi}{c^3} \nu_{lu}^3 \frac{g_l}{g_u} \frac{1}{n_l t}. \quad (7)$$

Hence it follows that the escape rate of L$\alpha$ photons from the $2p \to 1s$ line profile for the hydrogen atom due to the Sobolev mechanism of intrinsic photon escape per atom is greater in the outer layers, where the density of matter is lower. By contrast, since the two-photon $2s \to 1s$ decay rate per atom is constant, it follows that this decay will dominate in the inner subphotospheric zones of the expanding envelope. In these deep Lagrangian zones $A_{2p,1s}^{\text{eff}} n_{H^0,2p} < A_{2s,1s}^{2q} n_{H^0,2s}$. Since in our formulation of the problem we assume an $l$-equilibrium, i.e., $n_{H^0,2p}/n_{H^0,2s} = 3$, at $\beta_{L\alpha} \sim 4 \times 10^{-9}$ the escape rates from the profile due to the velocity gradient and the two-photon decay become equal.

### Atomic Data

We use a hydrogen model atom that consists of ten levels corresponding to the first ten principal quantum numbers. The atomic data for hydrogen were taken from the flexible atomic code (FAC) computations (Gu 2008) for a model atom with a fine structure. Since considering the fine structure of hydrogen is beyond the scope of this paper, the FAC data were "folded" into the so-called superlevels (Hubeny and Lanz 1995). Thus, we assume an $l$-equilibrium in the kinetic system described below, implying that at a fixed principal quantum number $n$ the populations of the fine-structure sublevels are proportional to their statistical weights.

The FAC results were successfully tested when compared with the data from other papers. For example, the energy levels and Einstein coefficients were compared with the data from the NIST atomic spectra database (Version 5). The photoionization cross sections were compared both with the Seaton approximation with the correction factor from Gould's formula (Pauldrach 1987, Eq. (A1)) and with the data from Karzas and Latter (1961). The electron-impact ionization rates were compared with the formula from Johnson (1972). The electron-impact excitation rates during collisions with electrons were computed in FAC in the distorted-wave approximation. They were compared with the results from Anderson et al. (2000, 2002) obtained by the R-matrix method and "folded" to superlevels. We also compared our data with those from Vrinceanu et al. (2014). In the electron-impact $n \to n'$ excitation rates we observed discrepancies in the weakly energetic transitions $\Delta n = 1$. For example, in the range of temperatures of interest to us for $n \geqslant 5$ the FAC data are occasionally overestimated by a factor of 10. However, precisely for high closely spaced levels, where the populations are already insignificant compared to the weakly excited levels, the electron-impact excitation and deexcitation rates become equal (Baklanov et al. 2013), and hence, the values of the rates themselves play a minor role. Therefore, the accuracy of the collisional processes between closely spaced levels is unimportant.

The two-photon $2s \to 1s$ decay rate was taken from Nussbaumer and Schmutz (1984).

*Kinetic Equations*

The continuity equation in Eulerian coordinates is

$$\frac{\partial \rho}{\partial t} = -\nabla(\rho v), \qquad (8)$$

where $\rho$ is the density of the envelope expanding with a velocity $v$. In the Lagrangian formalism in the comoving frame we obtain

$$\frac{D\rho}{Dt} = -\rho(\nabla \cdot v). \qquad (9)$$

If the photospheric phase is considered in the period of a free homologous expansion, then Eq. (9) will be simplified to

$$\frac{D\rho}{Dt} + \frac{3\rho}{t} = 0. \qquad (10)$$

We will restrict ourselves to the case of a SN envelope composed of one type of atoms. The subsequent general reasoning can also be easily extended to the case of atoms of different types. The total rate of transitions between a discrete level $i$ for an atom or ion with a degree of ionization z and the remaining levels will then be

$$\frac{Dn_{z,i}}{Dt} + \frac{3n_{z,i}}{t} = K_{z,i}(t), \qquad (11)$$

where $K_{z,i}$ is the function of time for level $i$ that is written as

$$K_{z,i}(t) \qquad (12)$$
$$= \sum_{u>i}[n_{z,u}A_{ui} + J_{iu}(n_{z,u}B_{ui} - n_{z,i}B_{iu})]$$
$$- \sum_{l<i}[n_{z,i}A_{il} + J_{li}(n_{z,i}B_{il} - n_{z,l}B_{li})]$$
$$+ n_e \sum_{k \neq i} n_{z,k}C_{ki} - n_e n_{z,i} \sum_{k \neq i} C_{ik}$$
$$- n_{z,i}\sum_r (P_{ir} + n_e C_{ir})$$
$$+ n_e \sum_r n_{z^+,r}(R_{ri} + n_e C_{ri})$$
$$+ \sum_m n_{z^-,m}(P_{mi} + n_e C_{mi})$$
$$- \sum_m n_{z,i}n_e(R_{im} + n_e C_{im}),$$
$$i = 1, 2 \ldots .$$

Here, $n_{z,i}$ is the population of level $i$ for an atom or ion with a degree of ionization z; the index $r$ is the level number for $z^+$; the index $m$ is the level number for $z^-$; $n_{z^-,m}$ and $n_{z^+,r}$ are the populations of the preceding and succeeding degrees of ionization for an atom or ion, which are nonzero only if they are possible; $n_e$ is the number density of free electrons; $A_{ki}$ and $B_{ki}$ are the Einstein coefficients for spontaneous and induced transitions; $C_{ki}$ and $C_{ik}$ are the collisional transition rates between levels; $J_{ik}$ is the mean intensity of radiation in the $i \to k$ transition averaged over the line profile (3); $P_{ir}$ and $P_{mi}$ are the photoionization coefficients; $R_{ri}$ and $R_{im}$ are the radiative recombination coefficients including the induced processes; $C_{ir}$ and $C_{mi}$ are the collisional ionization coefficients; $C_{ri}$ and $C_{im}$ are the three-body collisional recombination coefficients (Mihalas 1978; Hubeny and Mihalas 2014).

For the number density of free electrons we can write the equation

$$\frac{Dn_e}{Dt} + \frac{3n_e}{t} = K_e(t), \quad (13)$$

where $K_e$ is obtained from the charge conservation law

$$K_e(t) = \sum_z z \sum_i K_{z,i}(t). \quad (14)$$

Substituting (12) into this equation yields

$$K_e(t) = \sum_{z,i,r} \left[ n_{z,i}(P_{ir} + n_e C_{ir}) \right. \quad (15)$$
$$\left. - n_{z^+} n_e (R_{ri} + n_e C_{ri}) \right] = \text{Ion}(t) - \text{Rec}(t),$$

where $\text{Ion}(t)$ and $\text{Rec}(t)$ are the total ionization and recombination rates, respectively.

Basically, the system consisting of Eqs. (11) written for all ionization states and levels and (13) is already closed. However, we change one of the equations in (11) using the law of conservation of the number of particles:

$$\frac{Dn_{z,j}}{Dt} = -\frac{3N(t)}{t} - \sum_{z,i \neq j} \frac{Dn_{z,i}}{Dt}. \quad (16)$$

Here, $N(t) = \sum_{z,i} n_{z,i}$ is the total number density of particles, and level $j$ is chosen with a maximum population (Hubeny and Mihalas 2014, p. 282) to increase the numerical stability of our computation.

System (12) will be simplified in the case of hydrogen:

$$K_{H^0,i}(t) = \sum_{u>i} \left[ n_{H^0,u} A_{ui} \right. \quad (17)$$
$$\left. + J_{iu}(n_{H^0,u} B_{ui} - n_{H^0,i} B_{iu}) \right]$$
$$- \sum_{l<i} \left[ n_{H^0,i} A_{il} + J_{li}(n_{H^0,i} B_{il} - n_{H^0,l} B_{li}) \right]$$
$$+ n_e \sum_{k \neq i} n_{H^0,k} C_{ki} - n_e n_{H^0,i} \sum_{k \neq i} C_{ik}$$
$$- n_{H^0,i}(P_{ic} + n_e C_{ic}) + n_e n_{H^+}(R_{ci} + n_e C_{ci}),$$
$$i = 1, 2 \ldots,$$

$$K_{H^+}(t) = \sum_i n_{H^0,i}(P_{ic} + n_e C_{ic}) \quad (18)$$
$$- \sum_i n_{H^+} n_e (R_{ci} + n_e C_{ci}).$$

Here, $P_{ic}$ is the total photoionization coefficient, $R_{ci}$ is the total radiative recombination coefficient including the induced processes, $C_{ic}$ is the total collisional ionization coefficient, and $C_{ci}$ is the total three-body collisional recombination coefficient. The two-photon $2s \to 1s$ decay should also be added. Since an $l$-equilibrium is assumed in this paper, $n_{H^0,2s} = n_{H^0,2}/4$ and $n_{H^0,2p} = 3/4 n_{H^0,2}$, where $n_{H^0}$ is the neutral hydrogen number density. In (12) for the second hydrogen level we should then subtract $A^{2q}_{2s,1s} n_{H^0,2s}$ from the right-hand part, where $A^{2q}_{2s,1s}$ is the two-photon decay probability. The reverse $1s \to 2s$ transition (two-photon absorption) rate is much lower than the $2s \to 1s$ rate, and we disregard this process. Indeed, $A^{2q}_{1s,2s} n^*_{H^0,1s} = A^{2q}_{2s,1s} n^*_{H^0,2s}$ in the case of thermodynamic equilibrium. Our calculations show that $n_{H^0,2s} \gg n^*_{H^0,2s}$, while $n_{H^0,1s} < n^*_{H^0,1s}$. Below, alluding to (11), we will imply allowance for the two-photon decay.

*Initial Conditions*

The so-called nebular approximation that is widely used in Monte Carlo SN simulations is well known (Mazzali and Lucy 1993; Lucy 1999; Kerzendorf and Sim 2014). In this approximation the radiative processes are assumed to dominate over the collisional ones. Of course, in our complete calculation we will not use this assumption, but it is convenient to illustrate an interesting property of our system. Following Kerner (1972), it can be shown that in the case of switched-off collisional processes the system of equations (11), (13), (16) is reduced to a vector equation:

$$\dot{\mathbf{n}} = \mathbf{A} \cdot \mathbf{n} + \mathbf{n} \cdot \mathbf{B} \cdot \mathbf{n}, \quad (19)$$

where $\mathbf{n}$ is the vector of unknown level populations, while $\mathbf{A} = \mathbf{A}(t)$ and $\mathbf{B} = \mathbf{B}(t)$ are the second- and third-rank tensors, respectively, containing the rates of radiative processes dependent only on time. This is a Riccati-type equation (Boyce and DiPrima 1986, p. 87) that is commonly encountered in the optimal control theory (William and Levine 1996). It is well known that if there is some particular solution of this equation $\mathbf{n}_1$, then a general solution of (19) can be obtained by the Euler (1760) substitution

$$\mathbf{n} = \mathbf{n}_1 + \mathbf{m}^{-1}, \quad (20)$$

where $\mathbf{m}$ is a new unknown that is a solution of the linear vector ordinary first-order differential equation derived by directly substituting (20) into (19). It turns out that $\mathbf{m}$ is a rapidly increasing function of time in magnitude, and all of the general solutions converge to one particular solution $\mathbf{n}_1$. Thus, the solutions

"forget" the initial conditions and reach some limiting regime. Our calculations show that the complete kinetic system with collisions behaves in the same way. All of the aforesaid allows the initial conditions to be chosen more freely, because far from the initial time we can trust the solution irrespective of this choice.

Nevertheless, the initial conditions for the populations of all levels and the electron number density for our time-dependent computation are calculated in the nebular approximation (Mazzali and Lucy 1993; Lucy 1999) for each individual Lagrangian zone at the level of the thermalization depth. The thermalization depth corresponds to the subphotospheric regions of the envelope where the matter temperature coincides with the radiation temperature and is taken from the `STELLA` modeling.

*Steady-State Approximation*

Let us now give a rigorous definition of the steady-state approximation. The solution of the nonlinear algebraic system written for all ionization stages z,

$$K_{z,i}(t) = 0, i = 1, 2, \ldots, \quad (21)$$
$$K_e(t) = 0,$$

is called the steady-state approximation of solving the complete kinetic system (11), (13), (16). The key question is whether the solutions of these systems coincide and, if not, why.

Here, it will be convenient to introduce the following definition: we will call the solution of the complete time-dependent system (11), (13), (16) $n^{\text{td}}$ (time-dependent) and the solution of the algebraic system (21) $n^{\text{ss}}$ (steady-state).

Imagine that the envelope ceased to expand at some time $t_1$ after SN explosion and all other characteristics, such as the matter temperature and the photospheric radiation intensity, are fixed. The population kinetics in such an envelope can be described by the system

$$\frac{D\tilde{n}_{z,i}}{D\tilde{t}} = K_{z,i}(\tilde{t})\big|_{t=t_1}, \quad (22)$$
$$i = 1, 2, \ldots, j-1, j+1, \ldots,$$
$$\frac{D\tilde{n}_e}{D\tilde{t}} = K_e(\tilde{t})\big|_{t=t_1},$$
$$\frac{D\tilde{n}_{z,j}}{D\tilde{t}} = -\sum_{z,i \neq j} K_{z,i}(\tilde{t})\big|_{t=t_1}.$$

On the right-hand sides of system (22) all coefficients are taken at the time $t_1$ of interest to us, while the unknowns $\tilde{n}_{z,i}(\tilde{t})$ and $\tilde{n}_e(\tilde{t})$ are already functions of new time $\tilde{t} \geqslant t_1$.

Obviously, solution (21) at $t_1$ is a particular solution of (22). Consequently, reasoning in the same way as for Eq. (20), we find that the limiting solutions of the integro-differential system of equations (22) are another way of seeking the steady-state approximation for (11), (13), (16), and the limiting values of $\tilde{n}(\tilde{t})$ when $\tilde{t} \to \infty$ again do not depend on the initial conditions.

If, nevertheless, the number densities from $n^{\text{td}}(t_1)$ are taken as the initial conditions, then system (22) will also have a different physical meaning. It will show how the level populations will relax with time to their steady-state values if any microscopic changes in the plasma are "switched off." Therefore, the coefficients of the system are taken to be constant.

Interestingly, since the time derivative of the particular solution is zero (it is a steady-state one), it follows from substitution (20) in this case that $\dot{\tilde{\mathbf{n}}} = \dot{\mathbf{m}}^{-1}$. Since $\mathbf{m}$ is a rapidly increasing function in magnitude, we find that a maximum of the magnitude of $\dot{\tilde{\mathbf{n}}}$ is observed at the very beginning of the relaxation process. It will then become increasingly slow with time $\tilde{t}$.

Equation (22) represented in form (19) contains the tensors $\mathbf{A}(t_1)$ and $\mathbf{B}(t_1)$ that are no longer functions of time; therefore, this vector equation is solved in quadratures (Derevenskii 2008). For example, the solution of the simplest "level + continuum" system follows from the initial conditions to the steady state in hyperbolic tangent of time. This means that the solution of system (22) $\tilde{n}(\tilde{t})$ for $\tilde{t} \to \infty$ will reach the steady-state solution $n^{\text{ss}}$ formally infinitely long. Of course, when all quantities are considered with some specified accuracy, the process takes a finite time that may be called the relaxation time of $\tilde{n}$.

*Method of Calculation*

Having obtained the thermodynamic and hydrodynamic parameters of the SN envelope divided into 100 Lagrangian mass zones from the `STELLA` computations, we simultaneously solve the kinetic system (11), (13), (16) and the transfer equation in lines (3) in each of these zones independently in the `LEVELS` package using a high-order implicit method of prediction and correction with an automatic choice of the order and time step proposed by Gear (1971).

As the envelope expands, the photosphere moves along the Lagrangian coordinate toward the stellar center. The outer boundary of the thermalization region, where the matter and radiation temperatures are close, also moves inward. For each specific Lagrangian zone the time-dependent computation begins at the instant this zone "escapes" from the thermalization region.

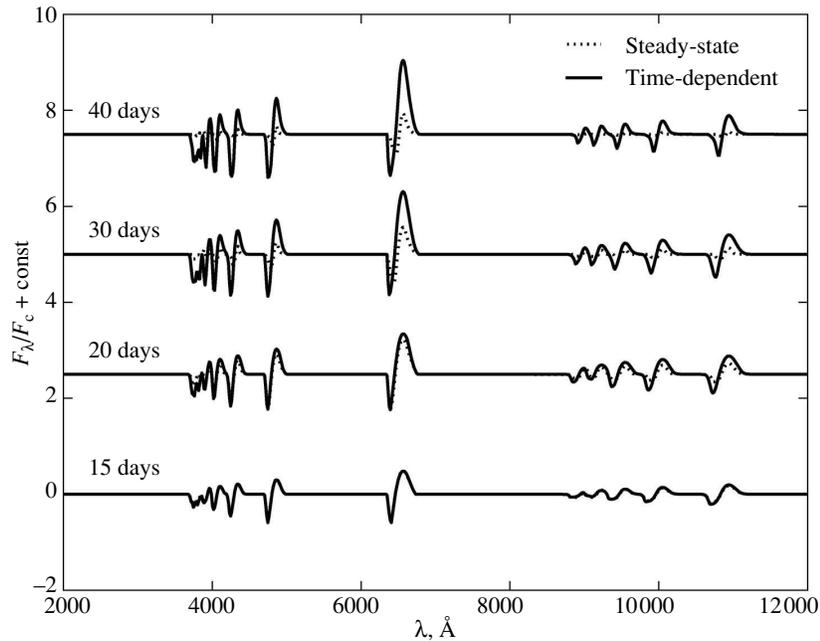

**Fig. 2.** Normalized spectra of the SN 1999em model for four times: 15, 20, 30, and 40 days after explosion, computed for the steady-state (dashed curve) and time-dependent (solid curve) cases.

The steady-state solutions are sought for pre-specified times by two different methods: as the solution of both system (21) and system (22).

For system (21) we apply Powell's hybrid method (Rabinowitz 1970, pp. 87–114) of solving nonlinear equations. For the stiff system of ordinary differential equations (22) we apply the implicit method by Gear (1971). In the steady-state approximation at the time under consideration the equations are solved only for those Lagrangian zones that are above the photosphere at this time.

The spectra are constructed in the Sobolev approximation by integrating the emitted flux for a given frequency over the surface of equal radial velocities (Mihalas 1978; Hubeny and Mihalas 2014).

## ILLUSTRATION OF THE TIME-DEPENDENT IONIZATION EFFECT

Figure 2 presents the normalized spectra of the SN 1999em model for four times: 15, 20, 30, and 40 days after explosion, computed for the steady-state and time-dependent cases. The atmosphere, i.e., the outer layers of the envelope lying above the photosphere, expands in SN 1999em homologously starting from day 10 (Baklanov et al. 2005; Utrobin 2007); therefore, our method of calculation is applicable. Note that the plateau phase in the light curve occurs on day 18–20 and, accordingly, the upper three spectra correspond to this phase. The time-dependent effect is pronounced both for the H$\alpha$ line and for the Balmer and Paschen series. Moreover, the time-dependent effect is enhanced with time starting from day 20.

The number density distribution of free electrons over the entire envelope for various times is presented in Fig. 3. In the time-dependent case, the envelope is seen to expand with a higher degree of ionization for all Lagrangian zones than in the steady-state approximation. The number density of free electrons is commonly said to experience "freeze-out" (Raizer 1959; Zel'dovich and Raizer 2008).

The system always tends to a kinetic equilibrium described by the steady-state approximation (21). For example, if at some time the level populations and the electron number density deviate from their steady-state values, then relaxation will return the system to these values. As has already been noted, if we imagine that the envelope ceased to expand at some time $t_1$ and all other characteristics, such as the matter temperature and the photospheric radiation intensity, were fixed, then the system would reach the current steady-state population $n^{\mathrm{ss}}(t_1)$ following the $\tilde{n}(\tilde{t})$ solution of (22), which is relaxation. If the characteristic time of change in the steady-state solution $n^{\mathrm{ss}}$ is much longer than the relaxation time of $\tilde{n}$ for any $t_1$, then the system will be able to "keep track" of this solution.

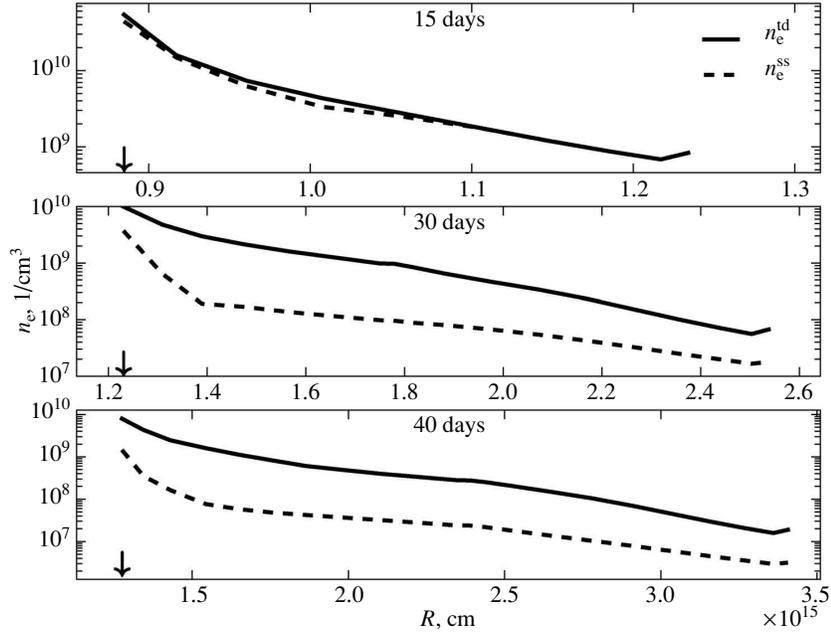

**Fig. 3.** Number density distribution of free electrons over the envelope as a function of radius for three times after explosion for the steady-state (dashed curve) and time-dependent (solid curve) cases. The arrow indicates the position of the photosphere.

Since the maximum of the function $|D\tilde{n}/D\tilde{t}|$ occurs at the initial time of the relaxation process $t_1$, if the inequality

$$\left|\frac{Dn^{\mathrm{ss}}}{Dt}\right|_{t=t_1} > \left|\frac{D\tilde{n}}{D\tilde{t}}\right|_{\tilde{t}=t_1} \quad (23)$$

holds, the condition for applicability of the steady-state approximation will be violated at $t_1$. If the inequality is valid for any $t_1$, then the system will never be able to be "adjusted" to the variable plasma parameters. Thus, condition (23) is sufficient for the breakdown of the steady-state approximation. On the other hand, however, it is clear that sufficiency is not equivalence, and the inverse inequality does not guarantee the applicability of the steady-state approximation.

As an example, consider the change of the electron number density in some Lagrangian zone close to the photosphere (Fig. 4). As the envelope expands, the photosphere moves along the Lagrangian coordinate toward the stellar center. The region under consideration "escapes" from beneath the photosphere on day 22 after explosion, when the plateau phase has already begun (Baklanov et al. 2005; Utrobin 2007). In Fig. 4 the dashed curve corresponds to the rate of change in the electron number density for the steady-state solution, while the solid one corresponds to the maxima of the relaxation rates. Condition (23) is violated almost everywhere and, accordingly, the time-dependent solution exhibits freeze-out (Fig. 5). In other words, the degree of ionization in the time-dependent case is higher than that in the steady-state one, and we observe the time-dependent effect.

Let us now consider two characteristic times: the so-called recombination time

$$t_{\mathrm{rec}} = \left|\frac{n_e}{\frac{D\tilde{n}_e}{D\tilde{t}}}\right|_{\tilde{t}=t_1} = \left|\frac{n_e}{K_e(\tilde{t})}\right|_{\tilde{t}=t_1} \quad (24)$$

(see, e.g., De et al. 2010a, Eq. (3)) and the characteristic time of change in the steady-state solution

$$t_{\mathrm{ss}} = \left|\frac{n_e}{\frac{Dn_e^{\mathrm{ss}}}{Dt}}\right|_{t=t_1}, \quad (25)$$

which corresponds to the characteristic time of change in macroscopic plasma parameters. If at some time $t_1$ the solutions $n_e^{ss}$ and $n_e^{td}$ are assumed to be still close, but inequality (23) holds, then inverting (23) written for the number density of free electrons and additionally multiplying it by $n_e(t_1)$, we will obtain a new sufficient condition for the breakdown of the steady-state approximation:

$$t_{\rm rec} > t_{\rm ss}. \qquad (26)$$

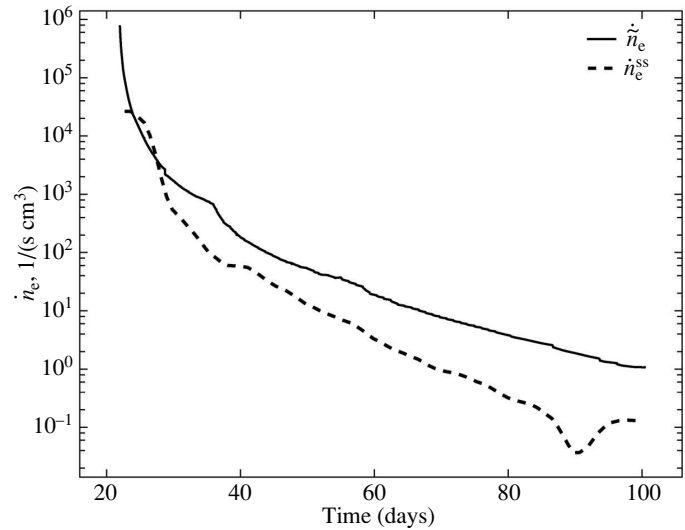

**Fig. 4.** Rate of change in the electron number density for the steady-state solution versus time in a Lagrangian layer in the envelope near $v \approx 5.8 \times 10^3$ km s$^{-1}$ (dashed curve). The maximum relaxation rate versus time in the same Lagrangian layer (solid curve).

However, it is important to note again that the condition inverse to (26) by no means guarantees the applicability of the steady-state approximation. The recombination time is basically a linear estimate of the true relaxation time, which can be much longer, because the transition to the steady state is a nonlinear process.

A similar situation with the erroneous zero estimate can be observed in the theory of burning according to the Arrhenius law (Frank-Kamenetskii 1939, 1987). In this case, the burning rate increases as the process develops, and the burning time estimated at the beginning turns out to be considerably longer than the actual one. By contrast, in our case, the relaxation rate slows down as the process proceeds, and the recombination time estimated at the initial time, on the contrary, can be shorter than the actual one.

To summarize, it can be noted that one should be careful when working with the linear estimate of the recombination time. An improvement in the applicability of the steady-state approximation does not necessarily follow from the decrease in the recombination time due to the inclusion of new channels in the kinetic system. It is best to use a comparison of the spectra in two types of calculations, the steady-state and time-dependent ones, to see the resulting effect. A difference in the strength of spectral lines is a reliable indicator of the intensity of the time-dependent effect (cf. Fig. 2).

Finally, it is important to note that the ordinary classical recombination time $t_{\rm rec}^{\rm class} = 1/(n_e \alpha_B)$ (Osterbrock and Ferland 2006, p. 22) is approximately equal to the recombination time (24) only in those regions and at those times where and when the photoionization rates are much lower than the recombination ones. This occurs at the late envelope expansion phases or in the outer layers, when the radiation is strongly diluted. In the inner subphotospheric layers or at the beginning of the plateau phase the following inequality holds:

$$t_{\rm rec} > t_{\rm rec}^{\rm class} \qquad (27)$$

(Utrobin and Chugai 2005b; Dessart and Hillier 2007).

In cosmology the fulfilment of inequality (27) is called the "protraction" of recombination (Kurt and Shakhvorostova 2014). Ionization freeze-out, i.e., the expansion of matter with a higher degree of ionization than that under equilibrium conditions, occurs due to the protraction.

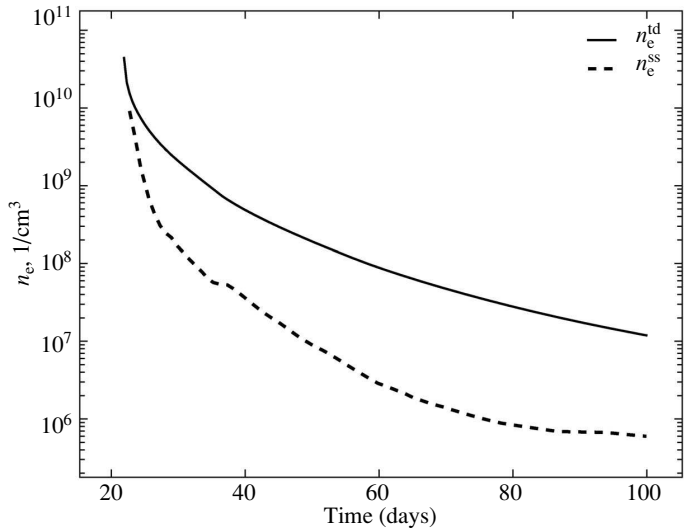

**Fig. 5.** Change of the electron number density in a Lagrangian layer in the envelope near $v \approx 5.8 \times 10^3$ km s$^{-1}$ with time in the steady-state (dashed curve) and time-dependent (solid curve) cases.

Zeldovich et al. (1968) and Peebles (1968) showed that under cosmological conditions the main processes leading to an increase in the number density of hydrogen atoms in the ground state are the two-photon decay of the $2s \to 1s$ level and the escape of L$\alpha$ photons from the line profile due to the Sobolev mechanism of intrinsic photon escape in the $2p \to 1s$ transition. The rates of these processes are less than or comparable to the rate of change in the steady-state solution.

This conclusion is basically confirmed in our calculations. Inequality (23) written for the ground hydrogen level holds for all zones of the envelope during the entire plateau phase. Hence, everywhere under conditions typical of SN 1999em the filling rate of the first level in the time-dependent description is always slower than that in the steady-state one. It is also important to note that the escape probability of resonant Lyman photons from the profile increases with level number (Hummer and Storey 1992; De et al. 2010b).

Our calculations show that the populations of levels with $n \geqslant 3$ are much smaller than that of the first excited level. Accordingly, the total effective population rate of the ground level due to the escape of resonant photons from the profile for these levels is much lower (by two or three orders of magnitude for deep subphotospheric layers) than the escape rate of L$\alpha$ photons. An increase in the number of levels in the model atom affects very weakly this total rate. Thus, in our case, we do not confirm the conclusions from De et al. (2010a) about the removal of the time-dependent effect as the number of levels increases. Since the first level in our formulation of the problem is in detailed balance with the continuum, the following channels are dominant and determine the population of the first level: the two-photon decay, the escape of L$\alpha$ photons, and, to a lesser degree, the collisional imbalance of the second level.

## WEAKENING OF THE TIME-DEPENDENT EFFECT

The previous section leads us to conclude that to weaken the time-dependent effect, the rates of the processes between levels 2 and 1 of the hydrogen atom need to be "accelerated." This implies an increase in the number of included channels between these levels. Such an increase can be achieved if we abandon the purely hydrogen chemical composition and take into account the contribution of additional "absorbers." There are two additional channels for an optically thick (in the L$\alpha$ line) SN envelope: the absorption of photons in flight in continuum (Hummer and Rybicki 1985; Chugai 1987) and the absorption in metal lines (a large number of Fe II and Cr II lines are in the vicinity of L$\alpha$) followed by cascade fragmentation to softer photons (Chugai 1998). In this case, the classical Sobolev photon escape probability $\beta_{L\alpha}$ can be approximately replaced in (3) by

$$\beta_{L\alpha}^{\text{eff}} \sim \beta_{L\alpha} + \beta_{L\alpha}^{C} + \beta_{L\alpha}^{L}, \qquad (28)$$

where $\beta_{L\alpha}^C$ and $\beta_{L\alpha}^L$ are the probabilities for the loss of a photon in flight when absorbed in continuum and metal lines, respectively.

Chugai (1998) showed that under typical conditions for SN II $\beta_{L\alpha}^L$ can exceed $4 \times 10^{-9}$, and, consequently, the two-photon decay rate ceases to dominate even in deep subphotospheric layers. At the nebular phase for type IIP supernovae Jerkstrand et al. (2012, 2011) estimated $\beta_{L\alpha}^L \approx 10^{-9}-10^{-8}$.

According to Chugai (1998), the probability for the loss of a photon in flight when absorbed in metal lines is given by the expression

$$\beta_{L\alpha}^L = \frac{k_z}{k_{L\alpha}} \sim \frac{\sum f_z n_z}{f_{L\alpha} n_1}, \qquad (29)$$

where $k_{L\alpha}$ is the absorption coefficient in the L$\alpha$ line, $k_z$ is the total absorption coefficient in the lines of metals capable of absorbing L$\alpha$, $f_z$ are the oscillator strengths for these lines, $f_{L\alpha}$ are the oscillator strength for the $2p \to 1s$ transition of the hydrogen atom, $n_1$ is the population of the ground hydrogen level, and $n_z$ is the population of the corresponding metal level. We will assume for our estimation that the metals are represented by Fe II ions and consider only the transitions from the lower $a^4D$ term with an excitation energy of $\sim 1$ eV. We will assume that all Fe II lines are very close to the center of the Voigt L$\alpha$ line profile. Substituting the oscillator strengths from the list of lines (on the red side of L$\alpha$) from Johansson and Jordan (1984, Table 1) into (29) and estimating the iron populations from the Boltzmann and Saha formulas at an abundance $X_{Fe} \sim 4 \times 10^{-4}$ in the range of matter temperatures $T_e \sim 3000-10\,000$ K, we find that $\beta_{L\alpha}^L \approx 10^{-9}-5 \times 10^{-8}$. It should be noted that at typical optical depths $\tau_{L\alpha} \sim 10^{10}$ the absorption in metal lines dominates over other channels, and we have $\beta_{L\alpha}^{eff} \sim \beta_{L\alpha}^L$.

Let us recalculate the previously obtained spectra (Fig. 2) by replacing the Sobolev probability $\beta$ in the $2p \to 1s$ transition with the maximum value that we estimated, $5 \times 10^{-8}$. The previous H$\alpha$ profiles and the new calculation are indicated, respectively, by the light-gray and black colors in Fig. 6. It can be seen that the introduced change affected the steady-state solution by increasing the line strength, while the line strength weakened but not that significantly in the time-dependent description. Thus, the presented estimates lead us to conclude that the processes under consideration do not cancel the time-dependent effect. Remarkably, if $\beta_{L\alpha}^{eff}$ is artificially increased by a factor of 10, then the H$\alpha$ profile for the time-dependent solution will weaken even more and will coincide with the steady-state one. Consequently, the abundance of iron (and other metals capable of absorbing L$\alpha$) in the SN envelope is an important factor affecting the intensity of the time-dependent effect.

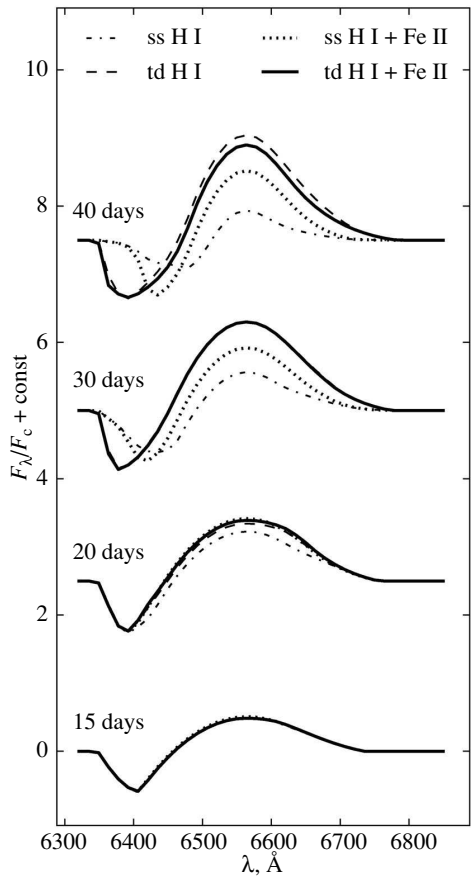

**Fig. 6.** Normalized H$\alpha$ profiles in the SN 1999em model for four times: 15, 20, 30, and 40 days (from bottom to top) after explosion. H I is the same as that in Fig. 2. The black dotted curve indicates the steady-state calculation with admixtures. The black solid curve indicates the time-dependent calculation with admixtures.

Yet another possibility should be pointed out. As in any plasma, collective electric fields are present in an expanding SN envelope. Such electric fields, being imposed on a hydrogen atom in the $2s$ state, reduce the lifetime of the atom, and it radiates an L$\alpha$ photon in the $2s \to 1s$ transition in a one-photon process. It is well known from the general theory (Bethe and Salpeter 1957) that when a uniform, constant, weak electric field of strength $E$ is imposed on a hydrogen atom in the metastable $2s$ state, the deexcitation time of this state changes and becomes

$t_{2p}(475\,\text{V/cm}/E)^2$, where $t_{2p}$ is the lifetime of the 2p state in the absence of any field. Even for fields with a strength of 0.05 V/cm the one-photon transition rate becomes equal to the two-photon one in the absence of any field. Note that this is still not enough to completely cancel the time-dependent effect. The question of the presence of possible electric fields in SN envelopes at the photospheric phase and their strength is beyond the scope of this paper.

## CONCLUSIONS

Using a simple model of SN 1999em as an example in the case of a purely hydrogen envelope, we demonstrated the time-dependent ionization effect by means of the STELLA and LEVELS software packages. We showed the intensity of the effect to increase with time and drew attention to the fact that allowance for additional metal admixtures could strongly affect the effect, while the addition of extra levels to the model hydrogen atom weakened the time-dependent effect to a lesser degree and never removed it completely, contrary to the assertion of De et al. (2010a).

The current realization of LEVELS is based on the already computed matter temperature profiles. Since this temperature depends on opacity and the latter, in turn, depends on level populations, it is extremely important to modify the software package for a self-consistent inclusion of the energy equation in it. The time-dependent effect in the energy equation (a deviation from the solution of the energy balance equation) can affect the final picture. The subsequent development of LEVELS will take into account the metals in the complete kinetic scheme. Nevertheless, even in the already existing form the LEVELS code allows one to rapidly investigate the role of various processes in the formation of SN spectra and to explain the discrepancies in the results of other groups of researchers concerned with similar problems.

It should be noted separately that for diffusive radiation we took the approximation of detailed balance for high-energy photons. In future we are planning to solve the time-dependent transfer equation in this frequency range. For continuum radiation we are planning to use multi-temperature parametrization of the radiation obtained in our STELLA computations. This will allow the difference between the atmospheric steady-state approximation and the time-dependent case to be studied more comprehensively.

## ACKNOWLEDGMENTS

We thank A. Andronova for the earliest version of the code, P. Baklanov, N. Shakhvorostova, A. Bykov, S. Sim, and E. Sorokina for useful discussions, and P. Baklanov for Fig. 1. This study was supported by the Russian Science Foundation (project no. 14-12-00203).

*Translated by V. Astakhov*